\documentclass[leqno,12pt,a4paper]{article}
\usepackage{setspace}
\usepackage{latexsym}
\usepackage{amsmath}
\usepackage{amssymb}
\usepackage{mathtools}
\usepackage{bbm}
\usepackage{amsthm}
\usepackage{siunitx} 
\DeclareSIUnit\time{\textit{Time}}
\usepackage[misc]{ifsym}

\usepackage[]{algorithm,algpseudocode}
\usepackage{comment}
\usepackage{graphicx}
\usepackage[colorlinks=true, allcolors=blue]{hyperref}
\usepackage{cancel}
\usepackage{mwe}
\usepackage{realhats}
\usepackage{amsbsy}
\usepackage{mathabx}
\usepackage{algorithm}
\usepackage{algpseudocode}
\usepackage{epsfig}
\usepackage{subcaption}
\usepackage{tabularx}
\usepackage{makecell}
\usepackage{array}
\usepackage{ulem}

\usepackage{import}
\usepackage{xcolor,colortbl}
\usepackage{caption}
\usepackage[utf8]{inputenc}
\usepackage{pgfgantt}
\usepackage{pdflscape}
\newcolumntype{L}[1]{>{\raggedright\arraybackslash}p{#1}} 
\newcolumntype{C}[1]{>{\centering\arraybackslash}p{#1}} 
\newcolumntype{R}[1]{>{\raggedleft\arraybackslash}p{#1}} 

\allowdisplaybreaks

\usepackage{tikz}
\usetikzlibrary{shapes,arrows,patterns,backgrounds,fit,arrows.meta}
\usetikzlibrary{spy,backgrounds}
\usepackage{pgfplots}

\usepackage{placeins}

\usepackage{todonotes}

\definecolor{LUHblue}{RGB}{0,80,155}
\definecolor{LUHblue40}{RGB}{153,185,216}
\definecolor{LUHblue20}{RGB}{204,220,235}
\definecolor{LUHblack40}{RGB}{153,153,153}
\definecolor{LUHblack20}{RGB}{204,204,204}
\definecolor{LUHgreen}{RGB}{200,211,23}
\definecolor{IKMgreen}{RGB}{175,218,0}

\makeatletter

\def\env@cases#1{%
  \let\@ifnextchar\new@ifnextchar
  \left\lbrace\def\arraystretch{1.2}%
  \array{@{}#1@{\quad}l@{}}}
\makeatother

\usepackage{booktabs}

\makeatletter
\let\oldtheequation\theequation
\renewcommand\tagform@[1]{\maketag@@@{\ignorespaces#1\unskip\@@italiccorr}}
\renewcommand\theequation{(\oldtheequation)}
\makeatother

\RequirePackage{relsize}
\newfont{\Sf}{cmssbx10 scaled 2074}

\usepackage{multirow}
\usepackage{lineno} 
\usepackage{array}
\usepackage{arydshln}
\author{}

\begin{document}
\newtheorem{thm}{Theorem}
\newtheorem{cor}[thm]{Corollary}
\newtheorem{lem}[thm]{Lemma}
\newtheorem{defi}[thm]{Definition}
\newtheorem{conj}[thm]{Conjecture}
\newtheorem{quest}[thm]{Question}

\newcommand{\argmax}{\operatornamewithlimits{argmax}}
\newcommand{\imax}{{i_\mathrm{m}}}

\newcommand{\boldface}[1]{\boldsymbol{#1}}
\newcommand{\bfa}{\boldface{a}}
\newcommand{\bfb}{\boldface{b}}
\newcommand{\bfc}{\boldface{c}}
\newcommand{\bfd}{\boldface{d}}
\newcommand{\bfe}{\boldface{e}}
\newcommand{\bff}{\boldface{f}}
\newcommand{\bfg}{\boldface{g}}
\newcommand{\bfh}{\boldface{h}}
\newcommand{\bfi}{\boldface{i}}
\newcommand{\bfj}{\boldface{j}}
\newcommand{\bfk}{\boldface{k}}
\newcommand{\bfl}{\boldface{l}}
\newcommand{\bfm}{\boldface{m}}
\newcommand{\bfn}{\boldface{n}}
\newcommand{\bfo}{\boldface{o}}
\newcommand{\bfp}{\boldface{p}}
\newcommand{\bfq}{\boldface{q}}
\newcommand{\bfr}{\boldface{r}}
\newcommand{\bfs}{\boldface{s}}
\newcommand{\bft}{\boldface{t}}
\newcommand{\bfu}{\boldface{u}}
\newcommand{\bfv}{\boldface{v}}
\newcommand{\bfw}{\boldface{w}}
\newcommand{\bfx}{\boldface{x}}
\newcommand{\bfy}{\boldface{y}}
\newcommand{\bfz}{\boldface{z}}
\newcommand{\bfA}{\boldface{A}}
\newcommand{\bfB}{\boldface{B}}
\newcommand{\bfC}{\boldface{C}}
\newcommand{\bfD}{\boldface{D}}
\newcommand{\bfE}{\boldface{E}}
\newcommand{\bfF}{\boldface{F}}
\newcommand{\bfG}{\boldface{G}}
\newcommand{\bfH}{\boldface{H}}
\newcommand{\bfI}{\boldface{I}}
\newcommand{\bfJ}{\boldface{J}}
\newcommand{\bfK}{\boldface{K}}
\newcommand{\bfL}{\boldface{L}}
\newcommand{\bfM}{\boldface{M}}
\newcommand{\bfN}{\boldface{N}}
\newcommand{\bfO}{\boldface{O}}
\newcommand{\bfP}{\boldface{P}}
\newcommand{\bfQ}{\boldface{Q}}
\newcommand{\bfR}{\boldface{R}}
\newcommand{\bfS}{\boldface{S}}
\newcommand{\bfT}{\boldface{T}}
\newcommand{\bfU}{\boldface{U}}
\newcommand{\bfV}{\boldface{V}}
\newcommand{\bfW}{\boldface{W}}
\newcommand{\bfX}{\boldface{X}}
\newcommand{\bfY}{\boldface{Y}}
\newcommand{\bfZ}{\boldface{Z}}
%
%
\newcommand{\bfalpha}{\boldsymbol{\alpha}}
\newcommand{\bfbeta}{\boldsymbol{\beta}}
\newcommand{\bfgamma}{\boldsymbol{\gamma}}
\newcommand{\bfdelta}{\boldsymbol{\delta}}
\newcommand{\bfvarepsilon}{\boldsymbol{\varepsilon}}
\newcommand{\bfzeta}{\boldsymbol{\zeta}}
\newcommand{\bfeta}{\boldsymbol{\eta}}
\newcommand{\bftheta}{\boldsymbol{\theta}}
\newcommand{\bfkappa}{\boldsymbol{\kappa}}
\newcommand{\bflambda}{\boldsymbol{\lambda}}
\newcommand{\bfmu}{\boldsymbol{\mu}}
\newcommand{\bfnu}{\boldsymbol{\nu}}
\newcommand{\bfpi}{\boldsymbol{\pi}}
\newcommand{\bfxi}{\boldsymbol{\xi}}
\newcommand{\bfrho}{\boldsymbol{\rho}}
\newcommand{\bfsigma}{\boldsymbol{\sigma}}
\newcommand{\bftau}{\boldsymbol{\tau}}
\newcommand{\bfphi}{\boldsymbol{\phi}}
\newcommand{\bfvarphi}{\boldsymbol{\varphi}}
\newcommand{\bfchi}{\boldsymbol{\chi}}
\newcommand{\bfomega}{\boldsymbol{\omega}}
\newcommand{\bfupsilon}{\boldsymbol{\upsilon}}
\newcommand{\bfeps}{\boldsymbol{\varepsilon}}
\newcommand{\bfpsi}{\boldsymbol{\psi}}
\newcommand{\bfGamma}{\boldsymbol{\Gamma}}
\newcommand{\bfDelta}{\boldsymbol{\Delta}}
\newcommand{\bfTheta}{\boldsymbol{\Theta}}
\newcommand{\bfLambda}{\boldsymbol{\Lambda}}
\newcommand{\bfPi}{\boldsymbol{\Pi}}
\newcommand{\bfXi}{\boldsymbol{\Xi}}
\newcommand{\bfSigma}{\boldsymbol{\Sigma}}
\newcommand{\bfPhi}{\boldsymbol{\Phi}}
\newcommand{\bfChi}{\boldsymbol{\Chi}}
\newcommand{\bfOmega}{\boldsymbol{\Omega}}
%
%
\newcommand{\calA}{\mathcal{A}}
\newcommand{\calB}{\mathcal{B}}
\newcommand{\calC}{\mathcal{C}}
\newcommand{\calD}{\mathcal{D}}
\newcommand{\calE}{\mathcal{E}}
\newcommand{\calF}{\mathcal{F}}
\newcommand{\calG}{\mathcal{G}}
\newcommand{\calH}{\mathcal{H}}
\newcommand{\calI}{\mathcal{I}}
\newcommand{\calJ}{\mathcal{J}}
\newcommand{\calK}{\mathcal{K}}
\newcommand{\calL}{\mathcal{L}}
\newcommand{\calM}{\mathcal{M}}
\newcommand{\calN}{\mathcal{N}}
\newcommand{\calO}{\mathcal{O}}
\newcommand{\calP}{\mathcal{P}}
\newcommand{\calQ}{\mathcal{Q}}
\newcommand{\calR}{\mathcal{R}}
\newcommand{\calS}{\mathcal{S}}
\newcommand{\calT}{\mathcal{T}}
\newcommand{\calU}{\mathcal{U}}
\newcommand{\calV}{\mathcal{V}}
\newcommand{\calW}{\mathcal{W}}
\newcommand{\calX}{\mathcal{X}}
\newcommand{\calY}{\mathcal{Y}}
\newcommand{\calZ}{\mathcal{Z}}
%
%

\newcommand{\dsA}{\mathbb{A}}
\newcommand{\dsB}{\mathbb{B}}
\newcommand{\dsC}{\mathbb{C}}
\newcommand{\dsD}{\mathbb{D}}
\newcommand{\dsE}{\mathbb{E}}
\newcommand{\dsF}{\mathbb{F}}
\newcommand{\dsG}{\mathbb{G}}
\newcommand{\dsH}{\mathbb{H}}
\newcommand{\dsI}{\mathbb{I}}
\newcommand{\dsJ}{\mathbb{J}}
\newcommand{\dsK}{\mathbb{K}}
\newcommand{\dsL}{\mathbb{L}}
\newcommand{\dsM}{\mathbb{M}}
\newcommand{\dsN}{\mathbb{N}}
\newcommand{\dsO}{\mathbb{O}}
\newcommand{\dsP}{\mathbb{P}}
\newcommand{\dsQ}{\mathbb{Q}}
\newcommand{\dsR}{\mathbb{R}}
\newcommand{\dsS}{\mathbb{S}}
\newcommand{\dsT}{\mathbb{T}}
\newcommand{\dsU}{\mathbb{U}}
\newcommand{\dsV}{\mathbb{V}}
\newcommand{\dsW}{\mathbb{W}}
\newcommand{\dsX}{\mathbb{X}}
\newcommand{\dsY}{\mathbb{Y}}
\newcommand{\dsZ}{\mathbb{Z}}
\newcommand{\be}{\begin{equation}}
\newcommand{\ee}{\end{equation}}
\newcommand{\bea}{\begin{eqnarray}}
\newcommand{\eea}{\end{eqnarray}}
\newcommand{\bes}{\begin{equation*}}
\newcommand{\ees}{\end{equation*}}
\newcommand{\beas}{\begin{eqnarray*}}
\newcommand{\eeas}{\end{eqnarray*}}
\newcommand{\D}{\displaystyle}
\newcommand{\tr}[1]{\mathrm{tr} \, #1}
\newcommand{\cof}[1]{\mathrm{cof}[#1]}
\newcommand{\inv}[1]{#1^{-1}}
\newcommand{\norm}[1]{\left\lVert#1\right\rVert}
\newcommand{\abs}[1]{\left\lvert#1\right\rvert}
\newcommand{\lap}{\mathop{}\!\mathbin\bigtriangleup}
\newcommand{\dd}{\ \mathrm{d}}

\newcommand{\re}[1]{{\color{blue}{#1}}}
\newcommand{\rev}[2]{\red{\textsuperscript{#1)}}\re{#2}}
\newcommand{\red}[1]{{\color{red}{#1}}}
\newcommand{\titem}{~~\llap{\textbullet}~~}

\newcommand{\Assemarg}[2]{{
  \underset{\mathsmaller{#1}}{\overset{\mathsmaller{#2}}%
  {\raisebox{-0.5ex}{\mbox{\Sf A}}}}\;%
  }}
\newcommand{\AssemargPJ}[2]{{
  \underset{#1}{\overset{\mathsmaller{#2}}%
  {\raisebox{-0.5ex}{\mbox{\Sf A}}}}\;%
  }}
  
\newcommand{\bb}[1]{\mathbb{#1}}
\newcommand{\vv}[1]{\pmb{#1}}
\newcommand{\bg}[1]{\pmb{#1}}
\newcommand{\nk}[1]{\mathrm{#1}}
\newcommand{\abl}[1]{\dot{#1}}
\newcommand{\pd}[2]{\frac{\partial{#1}}{\partial{#2}}}
\newcommand{\ablzeit}{\frac{\nk{d}}{\nk{d}t} }
\newcommand{\intok}[1]{\int\limits_{\Omega} #1 \, \nk{d}v }
\newcommand{\intdV}[1]{\int_{\Omega} #1 \ \mathrm{d} V}
\newcommand{\intda}[1]{\int_{\partial\omega} #1 \ \mathrm{d} a}
\newcommand{\intdv}[1]{\int_{\omega} #1 \ \mathrm{d} v}
\newcommand{\intdA}[1]{\int_{\partial\Omega} #1 \ \mathrm{d} A}
\newcommand{\conref}{^{(0)}}  

\usetikzlibrary{quotes,arrows.meta}
\tikzset{
  annotated cuboid/.pic={
    \tikzset{%
      every edge quotes/.append style={midway, auto},
      /cuboid/.cd,
      #1
    }
    \draw [every edge/.append style={pic actions, densely dashed, opacity=.5}, pic actions]
    (0,0,0) coordinate (o) -- ++(-\cubescale*\cubex,0,0) coordinate (a) -- ++(0,-\cubescale*\cubey,0) coordinate (b) edge coordinate [pos=1] (g) ++(0,0,-\cubescale*\cubez)  -- ++(\cubescale*\cubex,0,0) coordinate (c) -- cycle
    (o) -- ++(0,0,-\cubescale*\cubez) coordinate (d) -- ++(0,-\cubescale*\cubey,0) coordinate (e) edge (g) -- (c) -- cycle
    (o) -- (a) -- ++(0,0,-\cubescale*\cubez) coordinate (f) edge (g) -- (d) -- cycle;
  },
  /cuboid/.search also={/tikz},
  /cuboid/.cd,
  width/.store in=\cubex,
  height/.store in=\cubey,
  depth/.store in=\cubez,
  units/.store in=\cubeunits,
  scale/.store in=\cubescale,
  width=10,
  height=10,
  depth=10,
  units=cm,
  scale=.1,
}

%
%

\title{A continuum multi-species biofilm model with a novel interaction scheme}
\date{} 
\maketitle
{\large
\noindent{Felix Klempt}, Hendrik Geisler, Meisam Soleimani, Philipp Junker\\[0.5mm]
}
Leibniz University Hannover, Institute of Continuum Mechanics, Hannover, Germany\\[2mm]
{
Corresponding author:\\[0.5mm]
Felix Klempt, \color{LUHblue}{\Letter \hskip 1mm klempt@ikm.uni-hannover.de} 
}

\section*{Abstract}
Biofilms are complex structures which are inhabited by numerous amount of different species of microorganisms. Due to their ubiquity, they influence human life on an everyday basis. It is therefore important to understand the interactions between different biofilm components and reactions to outside conditions. For this purpose, mathematical models and \textit{in silico} experiments have proven themselves to be fundamental. In combination with \textit{in vitro} and \textit{in vivo} experiments, they can give more insights and focus researchers' attention, reducing costs in the process. In this work, a comprehensive multi-species continuum-based biofilm model is presented. This model is capable of replicating a variety of different biofilm interactions with an arbitrary number of species, while still being comprehensive to encourage usage by researchers less familiar with mathematical modeling. In addition to a nutrient source, antibiotic agents and their effect on the biofilm can also be depicted. The model is derived using Hamilton's principle of stationary action, ensuring thermodynamic consistency automatically. The results show good quantitative agreement with biofilm behavior.
\newline
\newline
\noindent\textbf{Keywords:} biofilm, multi-species, biofilm interaction, continuum model

\section{Introduction}\label{chapt:Introduction}
The necessity to model and simulate biofilm behavior stems from the ubiquity of microbial biofilms. It is estimated that half of the existing biomass are prokaryotes \cite{whitman1998prokaryotes,flemming2010biofilm}. Much of that mass is located in dense biofilm communities, containing hundreds of species, i.e., upwards of 500 different species in dental biofilms \cite{klapper2010mathematical, marsh2005dental}. In these multi-species systems, complex interactions and relationships between different species emerge \cite{picioreanu2004particle, james1995interspecies}. The first, theoretical interaction is Neutralism, if biofilms do not interact at all with each other. With a lack of enough resources, i.e., space, nutrients or other necessary compounds, different species of bacteria have to compete with each other. Often times, the biofilm is then dominated by one species, but other ones persist in the biofilm \cite{james1995interspecies}. If both species benefit from the presence of one another, it is called Protocooperation. This has been demonstrated often in biofilms \cite{james1995interspecies}. Other interactions are Allelopathy \cite{Rahman2015Mixed}, where one species produces chemicals, which in turn influence the others. This can lead to Ammensalism, where one species harms the other and Commensalism, where one species benefits from the other. These interactions lead to the stabilization of the biofilm \cite{james1995interspecies}. To control harmful and use beneficial biofilms, an understanding of the underlying principles is necessary, but the large number of interactions pose a challenge for \textit{in vitro} experiments  \cite{picioreanu2004advances}. Mathematical models can provide an environment where hypotheses can be tested rapidly and connections between different aspects can be drawn. Additionally, an \textit{in silico} experiment, i.e., a simulation, takes considerably less time than an \textit{in vitro} or an \textit{in vivo} experiment. Models based on mathematical descriptions have thus proven themselves to be a viable tool in analyzing complicated processes \cite{chaudhry1998review}. Prominent modeling approaches are Individual Based Modeling (IbM), cellular automaton (CA) and continuum-based modeling. IbM seems to be the natural approach since the biofilm itself consists of individual particles.
Behaviors like cell divisions and interactions between two cells can be modeled directly \textcolor{black}{as an interaction between two individual particles}. There exist several open-source software systems using this approach, such as 
BacSim \cite{Kreft1998BacSim}, Infobiotics Workbench \cite{Blakes2011Infobiotics},
iDynoMiCS \cite{lardon2011idynomics},
Simbiotics \cite{naylor2017simbiotics},
and NUFEB \cite{li2019nufeb}.
IbM models containing multiple species have been used to study the effects of EPS \cite{Kreft2001Effect} as well as detachment and sloughing \cite{Xavier2004Amodelling, Xavier2005Aframework} on multi species biofilm. In combination with the continuum approach, physical factors, like nutrient gradients, have been included \cite{picioreanu2004particle, Martin2015Assessing}.
To avoid the stochastic effects resulting from the \textcolor{black}{randomized placement of the individual particles in the} modeling approach of IbM, the same simulation \textcolor{black}{has to be repeated or it} can be combined with cellular automata rules, see \cite{Kreft2001Individual}.
Multi-species models based fully on cellular automata are, for example, \cite{Martin2017Anewmathematical, Tang2017Modeling, Tang2013Animproved}. Another way of modeling biofilm is the continuum approach. Here, the incorporation of governing laws from physics comes naturally. Thermodynamic consistency can be assured automatically when using the right modeling approach. The thermodynamics and mechanics of growth are discussed in detail by \cite{epstein2000thermomechanics} and \cite{lubarda2002mechanics}. Additionally, continuum-based models are deterministic, stochastic effects are thus not present and models do not need additional runs to average out random effects before conclusions can be drawn. Although the derivation of these models can be more complex, analytical studies can be performed to compare and characterize model results \cite{alpkvista2007multidimensional}. The complexity of these models makes them very predictive, but also hard to understand, especially for researchers not familiar with material modeling. Easier models can focus the attention on more important and relevant aspects of biofilm formation and growth, as well as species interaction. Multi-species continuum-based models in literature, which are suitable to model an arbitrary number of species, i.e., \cite{wanner1986multispecies, alpkvista2007multidimensional}, link the species via shared resources as space and nutrient available. This means that only competition between the species can be depicted. Models, which can depict Protocooperation, solve the problem of interaction with an additional interaction variable, which developes where one or both species are present and enhance the growth process of the other species \cite{soleimani2023numerical, Feng2021Modeling}. For a small number of species, i.e., dual-species systems, this is a suitable approach, but since for every interaction between two species an interaction variable is necessary, the number of internal variables and consequently the computational cost becomes unreasonable \textcolor{black}{when the number of species increases.}
\\
In this work, a continuum model is presented which provides a framework for integrating a theoretically arbitrary number of species into the biofilm. It is capable of displaying a wide variety of possible interactions from mutually inhibitory species to a symbiotic relationship. The integration of interaction is done via a constant interaction matrix. Thus, no additional internal variables need to be introduced. The model is derived from Hamilton's principle of stationary action, ensuring thermodynamic consistency. The energy density function is modeled to be comprehensive. The complexity of the model stems from the dissipation function, which is formulated as a function of the product of the internal variables. It thus interconnects the governing evolution equations leading to complex material behavior.
\\
The article is structured as follows. In \autoref{chapt:Methology}, the mathematical derivation of the model is presented. The necessary variables and assumptions are introduced and the governing equations are shown. In \autoref{chapt:Results}, a numerical study displaying the behavior of the models with different numbers of species, initial conditions and parameters is presented to show the versatility of the model. The last chapter, \autoref{chapt:Conclusion}, gives a short conclusion and an outlook for future investigations possible with this model.

\section{Mathematical description}\label{chapt:Methology}
In this work, a comprehensive model for biofilm growth and death is presented. An arbitrary number of species with interactions can be modeled.
The model is derived based on Hamilton's principle of stationary action. Hamilton's principle provides a physically sound way of material modelling, in particular, the automatic fulfillment of the first and second law of thermodynamics through the derivation of the model from a potential \cite{junker2021extended}. Thus, the biofilm model presented in this work is thermodynamically consistent. In addition, necessary constraints can be considered in Hamilton's principle simply by adding corresponding potentials. Hamilton's principle for a quasi-static and isothermal process is given by \cite{junker2021extended, klempt2024hamilton} as
\begin{equation}\label{extendHam}
    \mathcal{H} = \int\limits_{\tau}(\mathcal{G} + \mathcal{C} + \mathcal{D})\; \mathrm{dt} \rightarrow \underset{\boldsymbol{\xi}, \gamma}{\mathrm{stat}}.
\end{equation}
The Hamilton functional $\mathcal{H}$ consists of three parts: the total potential $\mathcal{G}$, the constraint functional $\mathcal{C}$ and the dissipation energy $\mathcal{D}$. Hamilton's principle states that $\mathcal{H}$ is stationary for all thermodynamic state variables. In this case, the thermodynamic state variables are the internal variables $\bfxi$ and the Lagrange multiplier $\gamma$. 
\\
In the following, the individual parts are described in more detail. The total potential $\mathcal{G}$ is given as
\begin{equation}
\mathcal{G} \coloneqq  \int\limits_{\Omega_0} {\Psi} \; \mathrm{dV} - \int\limits_{\Omega_0} \cancelto{0}{\bff^{\star}} \cdot \bfu \; \mathrm{dV} - \int\limits_{\partial \Omega_0} \cancelto{0}{\bft^{\star}} \cdot \bfu \; \mathrm{dA}
\end{equation} 
with the Helmholtz free energy $\Psi$, the density $\rho_0$, the body force $\bff^{\star}$, the traction force $\bft^{\star}$ and the displacements $\bfu$.
No external forces are applied to this biofilm model. Therefore, the external forces $\bff^{\star}$ and $\bft^{\star}$ are set to zero.
\\
The constraint functional 
\begin{equation}
\mathcal{C} \coloneqq \int\limits_{\Omega_0} \, \gamma \, (f(\bfxi)) \; \mathrm{dV},
\end{equation}
allows incorporation of the holonomic constraint $f(\bfxi)$ with the Lagrange parameter $\gamma$.
\\
The dissipated energy is given as
\begin{equation}
\mathcal{D} \coloneqq \int\limits_{\Omega_0} \mathrm{D}_{\mathrm{diss}} \; \mathrm{dV}.
\end{equation}
with the volume-specific dissipated potential $\textrm{D}_\textrm{diss}$.
\\
To arrive at a sensible description of a predictive model for biofilm growth and death, suitable model assumptions have to be made, which will be explained in the following.
\\
The set of internal variables $\bfxi$ is given as 
\begin{equation}
\bfxi = \begin{pmatrix} \bfphi \\ \bfpsi \end{pmatrix}.
\end{equation}
The volume fraction $\phi_i \; i \in \{1, 2, \cdots, n \}$, which indicates the volume covered by each of $n$ species of bacteria, as well as the empty space $\phi_0$, is listed in the vector
\begin{equation}
    \bfphi =
    \begin{pmatrix}
        \phi_0\\
        \phi_1\\
        \phi_2\\
        \vdots\\
        \phi_n
    \end{pmatrix} \; \mathrm{with} \; \phi_i \in [0,1], \; \forall i.
\end{equation}
In order to model the growth as well as the death of biofilm, a second state variable per biofilm species is necessary. The percentage of living bacteria of the $n$th species is given by
\begin{equation}
    \bfpsi =
    \begin{pmatrix}
        \psi_1\\
        \psi_2\\
        \vdots\\
        \psi_n
    \end{pmatrix}\; \mathrm{with} \; \psi_i \in [0,1], \; \forall i.
\end{equation}
Therefore, the volume covered by living bacteria of species $i$ is 
\begin{equation}
    \Bar{\phi}_i  \coloneqq \phi_i \, \psi_i \; \mathrm{with} \; \forall i. \label{eq:MultVars}
\end{equation} The amount of dead cells is consequently $\textcolor{black}{\tau_i} \coloneqq \phi_i \, (1-\psi_i) \; \mathrm{with} \; i \in  \{1, 2, \cdots, n \}$.\\
Since the total volume of all bacteria plus the empty space cannot exceed the available volume, the holonomic constraint
\begin{equation}
    f = \sum_{l=0}^n \phi_l - 1 = 0 \label{eq:Constraint}
\end{equation} 
has to be fulfilled at all times. \textcolor{black}{Since the model is formulated as a material point model, interpreting the empty space as physical voids or channels within the biofilm would be inappropriate. The empty space variable $\phi_0$ rather serves as a numerical auxiliary quantity used to enforce the holonomic volume constraint given by \autoref{eq:Constraint}}\\
The nutrients are modeled by a given variable $\mathrm{c}^{\star}$, and the antibiotics by the given variable $\alpha^{\star}$. Both are possibly time-dependent. 
Since living bacteria thrive with nutrients and the percentage of living bacteria reduces with antibiotics present, the energy density function is modeled as
\begin{equation}
    \Psi = - \frac{1}{2} \mathrm{c}^{\star} \; \Bar{\bfphi} \cdot \bfA \cdot \Bar{\bfphi} + \frac{1}{2} \alpha^{\star} \; \bfpsi \cdot \bfB \cdot \bfpsi.
\end{equation}
\textcolor{black}{The quadratic approach is chosen to allow for a linear dependency between nutrients and $\Bar{\phi}_i$ and antibiotics and $\psi_i$. Additionally, the quadratic approach ensures the convexity of the free energy density.}
The symmetric growth coefficient matrix
\begin{equation}
    \bfA =
\begin{pmatrix}
a_{11} & a_{12} & \cdots & a_{1n}\\
a_{12} & a_{22} & \cdots & a_{2n}\\
\vdots & \vdots  & \ddots & \vdots \\
a_{1n} & a_{2n} & \cdots & a_{nn}
\end{pmatrix}
\label{eq:AMatrix}
\end{equation}
and the antibiotic sensitivity matrix
\begin{equation}
    \bfB =
\begin{pmatrix}
b_{1} & 0 & \cdots & 0\\
0 & b_{2} & \cdots & 0\\
\vdots & \vdots  & \ddots & \vdots\\
0 & 0 & \cdots & b_{n}
\end{pmatrix}.
\label{eq:BMatrix}
\end{equation}
allow to specify the effect of nutrients and antibiotics on each species. Furthermore, the off-diagonal entries of the matrix $\bfA$ describe how the different species of bacteria interact with one another, i.e., if they promote or impair each others growth. \textcolor{black}{While in principle the entries of the matrices can have any form, i.e., be a function of any parameters or variables of the model, for simplicity reasons a zero-order Taylor series, i.e., a constant, is chosen to represent the interactions with nutrients and antibiotics. Hereby, a positive value indicates an equally favorable, a negative value an equally disadvantageous interaction.} The interaction between two species is consequently modeled with a single scalar with no need for additional state variables as in \cite{soleimani2023numerical}. This modeling choice vastly decreases simulation time.
\\
The dissipated energy is modeled as
\begin{equation}
    \mathrm{D_{diss}} \coloneqq \frac{\partial \Delta^s}{\partial \dot{\bfxi}} \cdot \bfxi.
\end{equation}
For more information on modeling assumptions concerning energy dissipation in the Hamiltonian framework, see \cite{junker2021extended}.
The dissipation function is modeled as 
\begin{equation}
    \Delta^s = \Delta^s(\Dot{\Bar{\bfphi}}, \Dot{\bfphi}) = \frac{1}{2} \; \Dot{\Bar{\bfphi}} \cdot \bfeta \cdot \Dot{\Bar{\bfphi}} + \frac{1}{2} \; \Dot{\bfphi} \cdot \bfeta \cdot \Dot{\bfphi}.
\end{equation} 
with the diagonal viscosity matrix
\begin{equation}
\bfeta =
\begin{pmatrix}
\eta_1 & 0 & \cdots & 0\\
0 & \eta_2 & \cdots & 0\\
\vdots & \vdots  & \ddots & \vdots\\
0 & 0 & \cdots & \eta_n
\end{pmatrix}.
\end{equation}
The dissipation function is the sum of two parts. The first part describes the dissipation due to the change in living biofilm. Due to the living biofilm being the product of the two state variables $\Bar{\phi_i} = \phi_i \psi_i$, the rates of the internal variables $\Dot{\phi}$ and $\Dot{\psi}$ are directly linked together. This allows for complicated, non-linear effects in the model. \textcolor{black}{Using only the first term of the dissipation function would lead to a rank-deficient system of differential equations. The second term depending only on the rate of the volume fraction of each biofilm leads to a full rank.}
The overall resulting behavior is rate-dependent.
\\
By evaluating Hamilton's principle, the strong form of the evolution equations are found as
\begin{align}
    \delta_{\phi} \mathcal{H} = 0 \; \forall \delta\phi \; \Leftrightarrow \; 0 &= -  c^{\star} \psi_i \left( a_{ii} \Bar{\phi}_i+ \sum_j^{n-1} a_{ij} \Bar{\phi}_j \right) + \eta_i (\Dot{\phi_i}\psi_i^2 +\Bar{\phi}_i \Dot{\psi}_i + \Dot{\phi}_i) + \gamma \label{eq:EvolEq1}\\    
    \delta_{\psi} \mathcal{H} = 0 \; \forall \delta\psi \; \Leftrightarrow \; 0 &= -c^{\star} \phi_i \left( a_{ii} \Bar{\phi}_i+ \sum_j^{n-1} a_{ij} \Bar{\phi}_j \right) +  \alpha^{\star} \psi_i b_i + \eta_i (\Dot{\psi_i}\phi_i^2 + \Bar{\phi}_i \Dot{\phi}_i) + \gamma \label{eq:EvolEq2}\\
    \delta_{\gamma} \mathcal{H} = 0 \; \forall \delta\gamma \; \Leftrightarrow \; 0 &= \sum_{l=0}^n \phi_l - 1 \label{eq:EvolLagrange}
\end{align}
The variable describing the empty space $\phi_0$ is described through the constraint in \autoref{eq:Constraint}.\\
The model was implemented using Wolfram Mathematica \cite{Mathematica} in an implicit framework on a material point. Therefore, the governing \autoref{eq:EvolEq1}, \autoref{eq:EvolEq2} and \autoref{eq:EvolLagrange} are solved using Newton's method at each time step.
The volume constraint \ref{eq:Constraint} is enforced by means of a Lagrange multiplicator $\gamma$.
To enforce $\phi_i, \psi_i \in [0,1]$, the barrier method is used with a penalty parameter $K_{p,i} =\eta_i \; 10^{-4}$ \cite{nesterov2018lectures}.

\section{Numerical results}\label{chapt:Results}
Several numerical experiments were performed to investigate the behavior of the model. First, a model with two species was chosen. This minimal model helps in understanding the interaction between different variables. For further investigation, a four species model was implemented. The four species model allows for a more thorough investigation of the model as well as showing more complex behaviors. The time step size $\Delta t$ is set constant to $\Delta t = \SI{e-4} \mathrm{Time}$ throughout the numerical experiments. \textcolor{black}{The unit of time can be chosen to represent an appropriate timescale for biofilm growth and has to be determined in subsequent \textit{in vitro} experiments.} 

\subsection{Two species}
For the two species model, five cases are investigated. They differ in the parameters listed in \autoref{tab:parameters-2-species}.
 
\begin{figure}[H]
    \centering
    \begin{subfigure}{0.45\textwidth}
        \centering
        \includegraphics[width=1\textwidth]{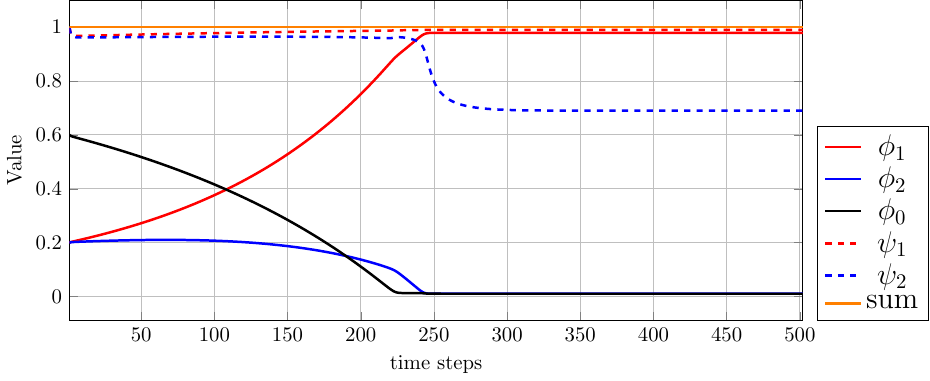}
        \caption{Volume fractions $\phi_i$ and percentage of living cells $\psi_i$ of test case 1}
        \label{subfig: Case1 2Species}
    \end{subfigure}
    \begin{subfigure}{0.45\textwidth}
        \centering
        \includegraphics[width=1\textwidth]{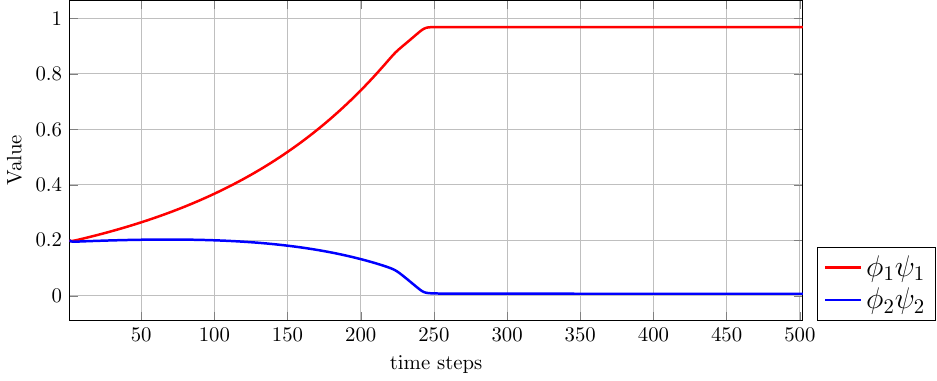}
        \caption{Amount of living cells $\Bar{\phi}_i$ of test case 1\newline}
        \label{subfig: Case1a 2Species}
    \end{subfigure}  
     \caption{Results of the test case with two species with different growth parameters. The species only interact through the space available and \textcolor{black}{don't enhance} each other's growth. Both species are fully resistant to antibiotics.}
     \label{fig: Case1 2Species}
\end{figure}

\begin{figure}[H]
    \centering
    \begin{subfigure}{0.45\textwidth}
        \centering
        \includegraphics[width=1\textwidth]{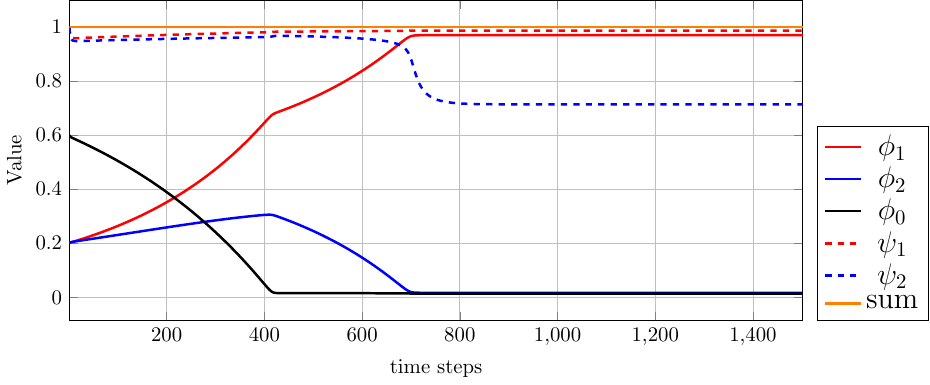}
        \caption{Volume fractions $\phi_i$ and percentage of living cells $\psi_i$ of test case 2}
        \label{subfig: Case2 2Species}
    \end{subfigure}
    \begin{subfigure}{0.45\textwidth}
        \centering
        \includegraphics[width=1\textwidth]{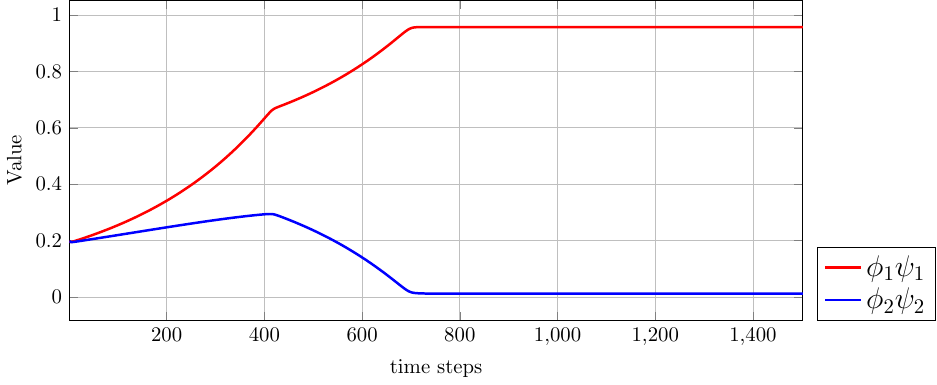}
        \caption{Amount of living cells $\Bar{\phi}_i$ of test case 2\newline}
        \label{subfig: Case2a 2Species}
    \end{subfigure} 
     \caption{Results of the test case with two species with different viscosities. The species only interact through the space available and do not enhance each other's growth. Both species are fully resistant to antibiotics.}
     \label{fig: Case2 2Species}
\end{figure}

\begin{centering}
    \begin{table*}[h!]
\begin{tabular}{llrrrrrr}\toprule
\textbf{variable} & \textbf{unit} & \textbf{Case 1} & \textbf{Case 2} & \textbf{Case 3} & \textbf{Case 4} & \textbf{Case 5}  & \textbf{Case 6}  \\ \midrule
${a}_{11}$ & [-] & 2 & 1 & 1 & 1 & 1 & 1 \\ \hdashline
${a}_{12}$ & [-] & 0 & 0 & 1 & 0 & 0 & -1\\ \hdashline
${a}_{22}$ & [-] & 1 & 1 & 1 & 1 & 1 & 1 \\ \midrule

${b}_{1}$ & [-] & 0 & 0 & 0 & 1 & 1 & 0  \\ \hdashline
${b}_{2}$ & [-] & 0 & 0 & 0 & 2 & 2 & 0 \\ \midrule

${\eta}_{1}$ & $[\si[per-mode=fraction]{\kg\per\meter\per \time }]$ & 1 & 1 & 1 & 1 & 1 & 1  \\ \hdashline
${\eta}_{2}$ & $[\si[per-mode=fraction]{\kg\per\meter\per\time}]$ & 1 & 2 & 2 & 2 & 2  & 2 \\ \midrule

{initial $\phi_{1}$} & [-] & 0.2 & 0.2 & 0.2 & 0.2 & 0.25 & 0.2\\ \hdashline
{initial $\phi_{2}$} & [-] & 0.2 & 0.2 & 0.2 & 0.3 & 0.3 & 0.2\\ \bottomrule

\end{tabular}
\caption{Values for the parameters of the simulations performed with two species present. The variables describing the nutrients $\mathrm{c}^{\star}$ and the antibiotics $\alpha^{\star}$ are left constant at $\mathrm{c}^{\star} = \SI[per-mode=fraction]{100}{\joule\per\meter\cubed}$ and $\alpha^{\star} = \SI[per-mode=fraction]{10}{\joule\per\meter\cubed}$.}
\label{tab:parameters-2-species}
\end{table*}
\end{centering}
\noindent The results from the first test case are depicted in \autoref{subfig: Case1 2Species} and \autoref{subfig: Case1a 2Species}. \autoref{subfig: Case1 2Species} shows the plot of the state variables $\bfphi$ and $\bfpsi$ over time. The amount of biofilm present from a species $\phi_i$ is given with a solid colored line. The empty space $\phi_0$ is shown in black. The dashed line represents the percentage of living bacteria of a given species $\psi_i$. In this first test case, the only interaction between both species is due to the limited amount of space available, since the non-diagonal terms of the growth parameter matrix $\bfA$ are set to zero. The diagonal entries of the growth parameters matrix are chosen to be $a_{11} > a_{22}$. This results in a faster growth rate of species 1. Once the empty space converges to zero, kinks in the graphs emerge and the volume fraction $\phi_2$ drops down rapidly. As soon as the volume fraction of species 2 reaches a value close to zero, its percentage of living cells $\psi_2$ drops down. Similar effects can be observed in the results of test case 2 in \autoref{subfig: Case2 2Species} and \autoref{subfig: Case2a 2Species}. In this test case, the viscosity of the species $\eta_i$ differs. A higher viscosity leads to a slower reaction of the biofilm. The simulation time is thus elongated to be $1500$ time steps. The aforementioned model behaviors from case 1 are also visible and even more prominent. While in case 1 the amount of species 2 $\phi_2$ is already at a low percentage of around $10\%$ and still declining before the empty space reaches zero. In case 2, $\phi_2$ is at roughly $30\%$ and ascending. The different amounts of species 2 present, result in different reaction from species 1. In the first case, the slope of the $\phi_1$-curve changes only slightly, while in case 2 a prominent kink is observable. The drop in the percentage of living bacteria of species 2 $\psi_2$ is comperable for both cases, as is the converged steady state solution. 

\begin{figure}[H]
    \centering
    \begin{subfigure}{0.45\textwidth}
        \centering
        \includegraphics[width=1\textwidth]{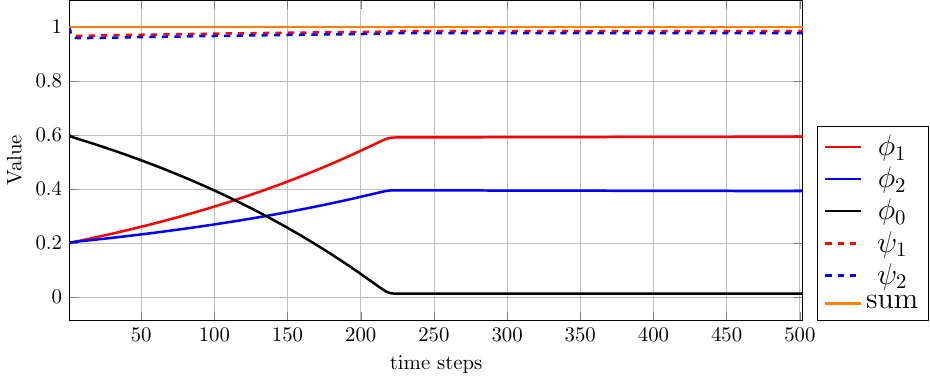}
        \caption{Volume fractions $\phi_i$ and percentage of living cells $\psi_i$ of test case 3}
        \label{subfig: Case3 2Species}
    \end{subfigure}
    \begin{subfigure}{0.45\textwidth}
        \centering
        \includegraphics[width=1\textwidth]{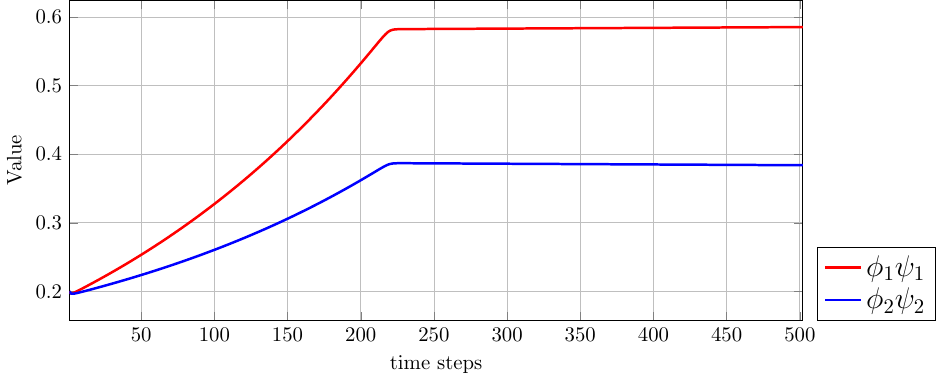}
        \caption{Amount of living cells $\Bar{\phi}_i$ of test case 3\newline}
        \label{subfig: Case3a 2Species}
     \end{subfigure}   
     \caption{Results of the test case with two species. The species interact and enhance each others growth. Both species are fully resistant to antibiotics.}
     \label{fig: Case3 2Species}
\end{figure}

\noindent When interactions between the species are enabled via $a_{ij} \neq 0 \; \mathrm{for} \; i \neq j$, as it is in test case 3, the survival of the other species is favorable for both species. An equilibrium between the two species emerges, where none gets extinguished. This can be observed in the result of test case 3 in  \autoref{subfig: Case3 2Species} and \autoref{subfig: Case3a 2Species}.

\begin{figure}[H]
    \centering
    \begin{subfigure}{0.45\textwidth}
        \centering
        \includegraphics[width=1\textwidth]{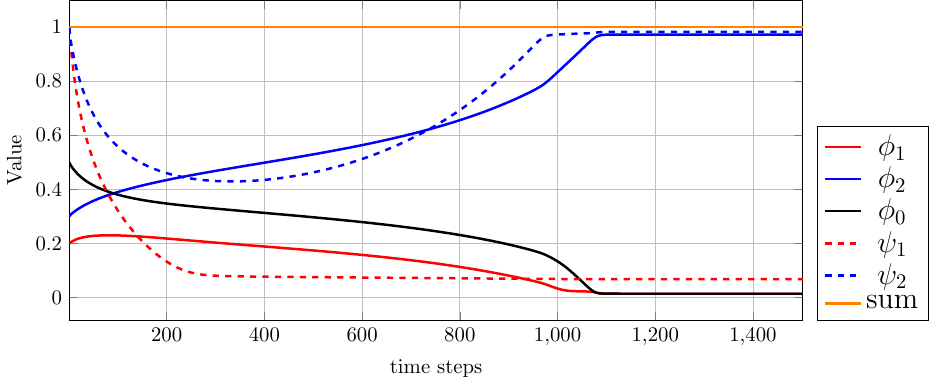}
        \caption{Volume fractions $\phi_i$ and percentage of living cells $\psi_i$ of test case 4}
        \label{subfig: Case4 2Species}
    \end{subfigure}
    \begin{subfigure}{0.45\textwidth}
        \centering
        \includegraphics[width=1\textwidth]{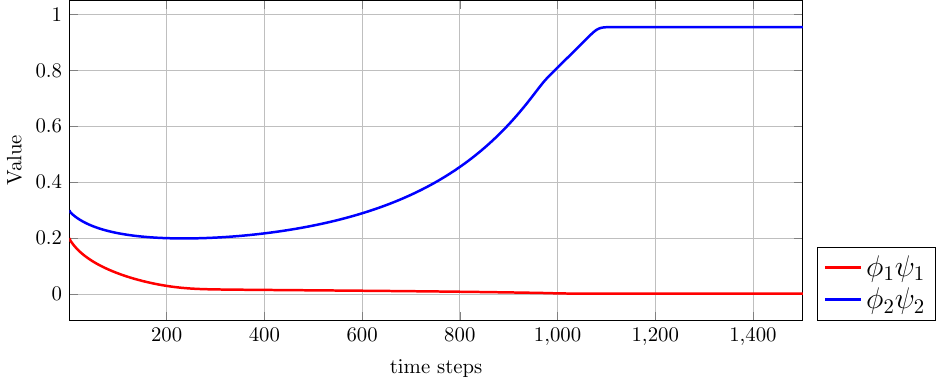}
        \caption{Amount of living cells $\Bar{\phi}_i$ of test case 4\newline}
        \label{subfig: Case4a 2Species}
    \end{subfigure}
     \caption{Results of the test cases with sensitivity of antibiotics. The initial volume fractions of species $\phi_i$ differ. Species 1 starts with an initial volume fraction of $\phi_1=0.2$, species 2 with $\phi_2=0.3$.}
     \label{fig: Case4 2Species}
\end{figure} 

\begin{figure}[H]
    \centering
   \begin{subfigure}{0.45\textwidth}
        \centering
        \includegraphics[width=1\textwidth]{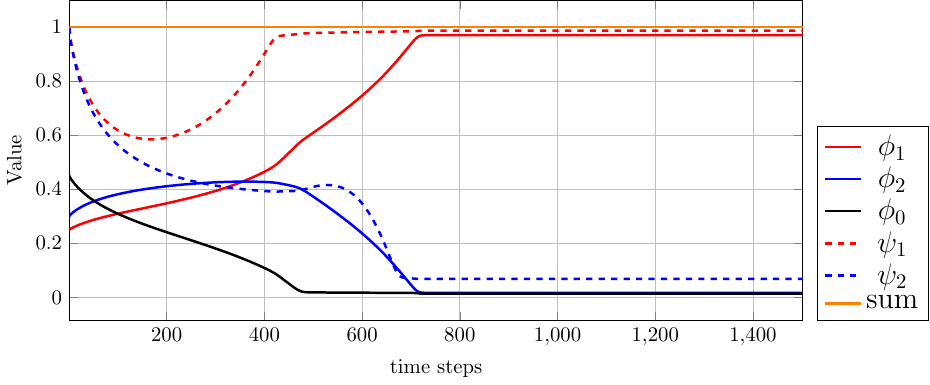}
        \caption{Volume fractions $\phi_i$ and percentage of living cells $\psi_i$ of test case 5}
        \label{subfig: Case5 2Species}
    \end{subfigure}
    \begin{subfigure}{0.45\textwidth}
        \centering
        \includegraphics[width=1\textwidth]{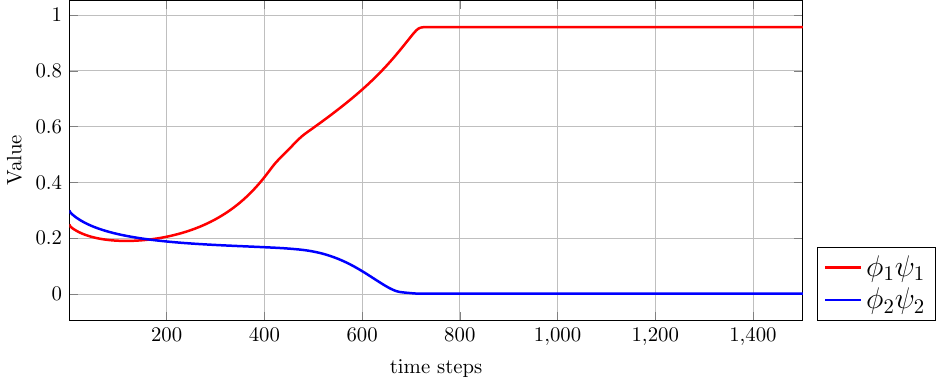}
        \caption{Amount of living cells $\Bar{\phi}_i$ of test case 5\newline}
        \label{subfig: Case5a 2Species}
    \end{subfigure}
     \caption{Results of the test cases with sensitivity of antibiotics. The initial volume fractions of species $\phi_i$ differ. Species 1 starts with an initial volume fraction of $\phi_1=0.2$, species 2 with $\phi_2=0.3$.}
     \label{fig: Case5 2Species}
\end{figure}
\begin{figure}[H]
    \centering
   \begin{subfigure}{0.45\textwidth}
        \centering
        \includegraphics[width=1\textwidth]{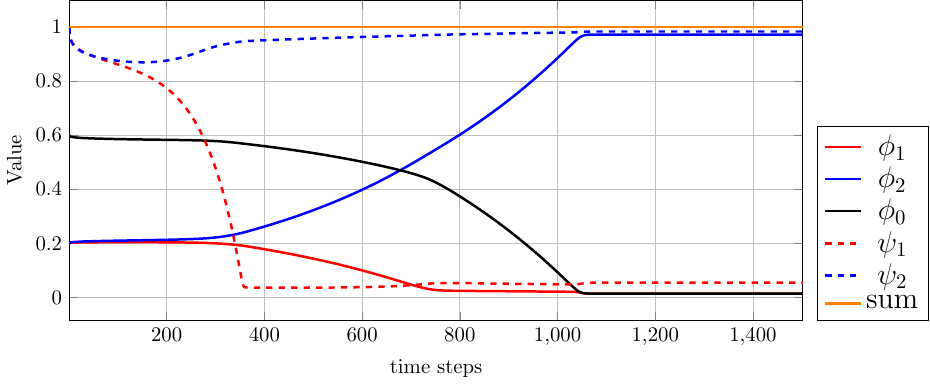}
        \caption{Volume fractions $\phi_i$ and percentage of living cells $\psi_i$ of test case 6}
        \label{subfig: Case6 2Species}
    \end{subfigure}
    \begin{subfigure}{0.45\textwidth}
        \centering
        \includegraphics[width=1\textwidth]{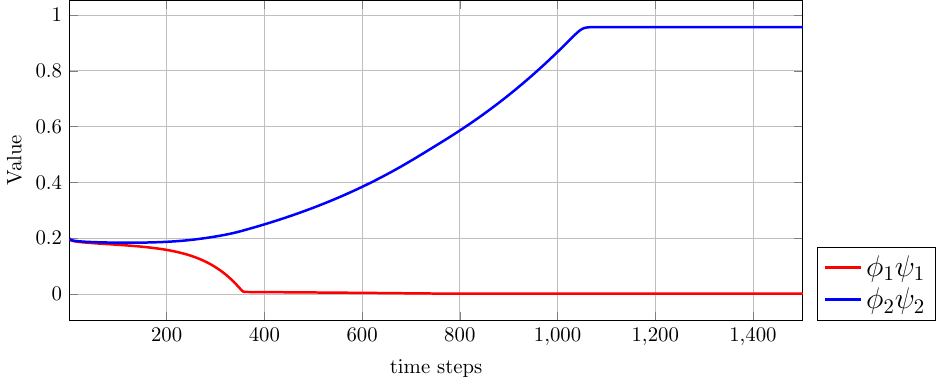}
        \caption{Amount of living cells $\Bar{\phi}_i$ of test case 6\newline}
        \label{subfig: Case6a 2Species}
    \end{subfigure}
     \caption{Results of the test cases with negative entry in the growth coefficient matrix.}
     \label{fig: Case6 2Species}
\end{figure}

 
\noindent For antibiotics sensitive biofilms, longer simulation times are necessary as well. It can be observed in \autoref{subfig: Case4 2Species} and \autoref{subfig: Case5 2Species}, that the percentage of living species $\psi_i$ drops initially. Although the sensitivity for antibiotics of species 2 is higher than that of species 1, due to the higher initial condition, species 2 can prevail over species 1 in the fourth test case. If the difference of initial biofilm is closer, as it is in test case 5, it is not enough and species 1 prevails instead. This can be seen in \autoref{subfig: Case5 2Species}. The amount of living bacteria of each species $\Bar{\phi_i} = \phi_i \psi_i$ can be observed for both cases in \autoref{subfig: Case4a 2Species} and \autoref{subfig: Case5a 2Species}.
\textcolor{black}{In all cases above, the interaction between different species of biofilm is modeled with a positive integer to model cooperative behavior. To model competition, a negative value for the off-diagonal entries of the interaction matrix $\mathbf{A}$ can be used. The results of this simulation are shown in \autoref{fig: Case6 2Species}. It can be observed, that the slower reacting species 2, i.e., the species with a higher viscosity $\eta_i$, comes out dominant. While a faster reaction is advantageous while reacting to nutrients, it can be detrimental when reacting negatively to another species. Early in the simulation, the percentage of living bacteria of species 1 $\psi_1$ drops of faster than the the living percentage of species 2 $\psi_2$. This leads to a difference in volume growth, where the species 2 with more living cells flourishes. The drop off in percentage of living cells is different from the experiment, where no interactions is present (c.f. \autoref{fig: Case1 2Species}). There is no negative part in the driving force in these simulation, thus the percentage of living cells $\psi_i$ stays at a relatively high level.}

\noindent\textcolor{black}{Additional simulation with varying viscosities $\eta_i$ were performed. The results are listed in \autoref{appendix:A}.}

\subsection{Four Species}
In theory, an arbitrary number of species can be inserted into the model. For this study, a biofilm consisting of four species is considered. The growth parameter matrix $\bfA$ stays the same throughout all numerical experiments in this chapter, as do the viscosities $\eta_i$, only the antibiotic sensitivity parameters $b_i$ as well as the nutrient concentration $c^{\star}$ and the concentration of antibiotics $\alpha^{\star}$ are changed. The values for the symmetric growth coefficient matrix $\bfA$ and the diagonal viscosity matrix $\bfeta$ are 
\begin{equation}
    \bfA = \frac{1}{2}
\begin{pmatrix}
1 & 5 & 5 & 5\\
5 & 1 & 3 & 3\\
\textcolor{black}{5} & 3 & 1 & 2 \\
5 & 3 & 2 & 1
\end{pmatrix}
\label{eq:4-species-AMatrix}
\end{equation}
and
\begin{equation}
\bfeta =
\begin{pmatrix}
0.8 & 0 & 0 & 0\\
0 & 1.0 & 0 & 0\\
0 & 0 & 1.5 & 0 \\
0 & 0 & 0 & 2.0
\end{pmatrix} \si[per-mode=fraction]{\kg\per\meter\per\time},
\label{eq:4-species-eta}
\end{equation}
respectively. The values for the remaining parameters of the four cases investigated are listed in \autoref{tab:parameters-4-species}.
 
\begin{centering}
    \begin{table}[h!]
\begin{tabular}{@{}llrrrr@{}}\toprule
\textbf{variable} & \textbf{unit} &  \textbf{Case 1} & \textbf{Case 2} & \textbf{Case 3} & \textbf{Case 4} \\ \midrule

${b}_{1}$ & [-] & 0.4 & 0.4 & 0.4 & 10\\ \hdashline
${b}_{2}$ & [-] & 0.3 & 0.3 & 0.3 & 2\\ \hdashline
${b}_{3}$ & [-] & 0.2 & 0.2 & 0.2 & 1\\ \hdashline
${b}_{4}$ & [-] & 0.1 & 0.1 & 0.1 & 0.01 \\ \midrule


{initial $\phi_{1}$} & [-] & 0.02 & 0.02 & 0.02 & 0.02\\ \hdashline
{initial $\phi_{2}$} & [-] & 0.02 & 0.02 & 0.02 & 0.02\\ \hdashline
{initial $\phi_{3}$} & [-] & 0.02 & 0.02 & 0.02 & 0.02\\ \hdashline
{initial $\phi_{4}$} & [-] & 0.02 & 0.2  & 0.02 & 0.02\\ \bottomrule

{nutrients $c^{\star}$} & $[\si[per-mode=fraction]{\joule\per\meter\cubed}]$ & 100 & 100 & $50 + 50 \; \mathrm{Sin}(500 t)$ & 100\\ \hdashline
{antibiotics $\alpha^{\star}$} & $[\si[per-mode=fraction]{\joule\per\meter\cubed}]$ & 10 & 10 & \textcolor{blue}{10} & $\mathrm{Piecewise} \; \{0,\{100,t>500\}\}$ \\ \bottomrule

\end{tabular}
\caption{Values for the parameters of the simulations performed with four species present.}
\label{tab:parameters-4-species}
\end{table}
\end{centering}

\begin{figure}[H]
    \centering
    \begin{subfigure}{0.45\textwidth}
        \centering
        \includegraphics[width=1\textwidth]{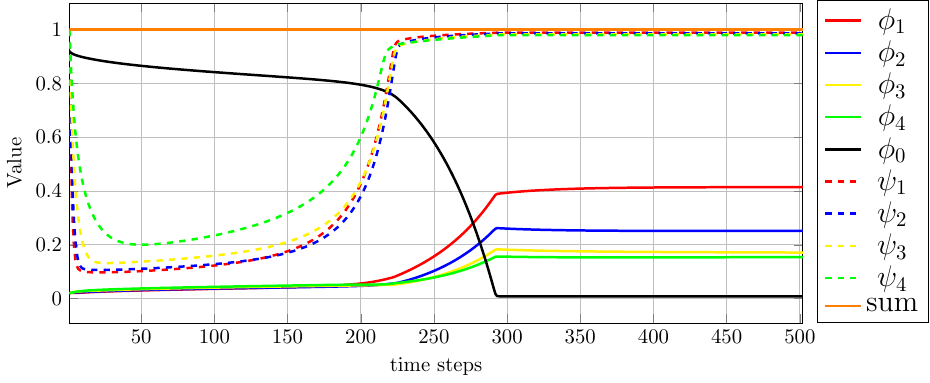}
        \caption{Volume fractions $\phi_i$ and percentage of living cells $\psi_i$ of test case 1}
        \label{subfig: Case1 4Species}
    \end{subfigure}
    \begin{subfigure}{0.45\textwidth}
        \centering
        \includegraphics[width=1\textwidth]{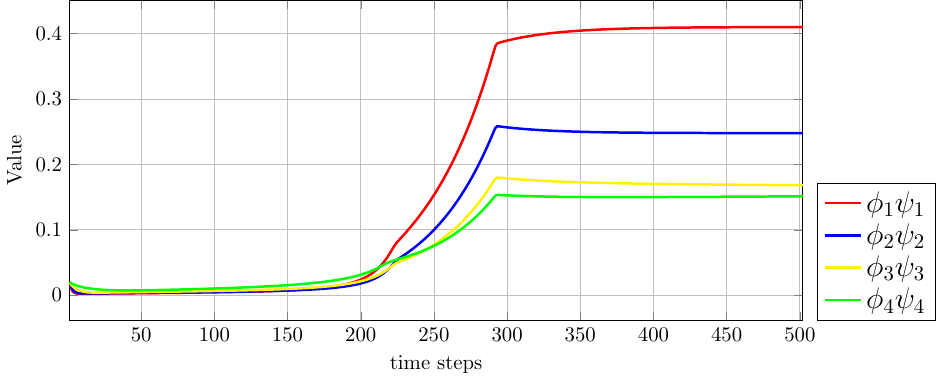}
        \caption{Amount of living cells $\Bar{\phi}_i$ of test case 1\newline}
        \label{subfig: Case1a 4Species}
    \end{subfigure}
     \caption{Result of the first test case with four species.}
     \label{fig: Case1 4Species}
\end{figure} 

\begin{figure}[H]
    \centering
   \begin{subfigure}{0.45\textwidth}
        \centering
        \includegraphics[width=1\textwidth]{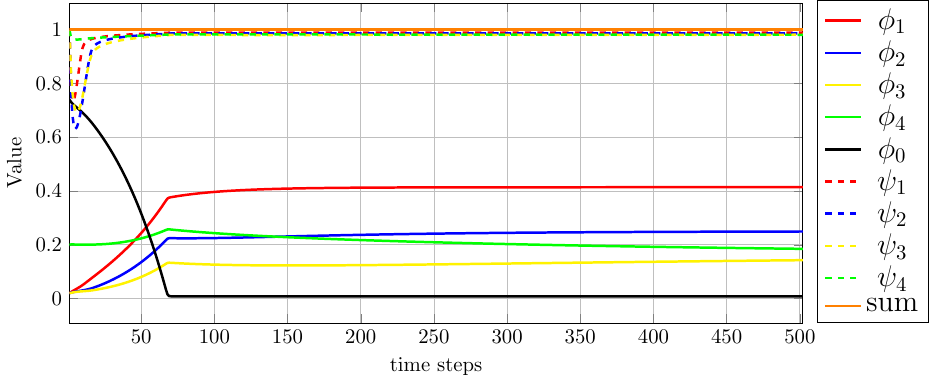}
        \caption{Volume fractions $\phi_i$ and percentage of living cells $\psi_i$ of test case 2}
        \label{subfig: Case2 4Species}
    \end{subfigure}
    \begin{subfigure}{0.45\textwidth}
        \centering
        \includegraphics[width=1\textwidth]{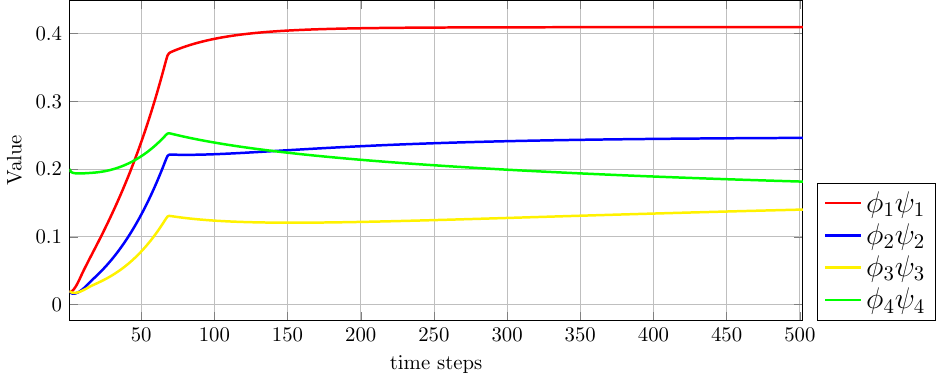}
        \caption{Amount of living cells $\Bar{\phi}_i$ of test case 2\newline}
        \label{subfig: Case2a 4Species}
    \end{subfigure}
     \caption{Result of the second test case with four species. The initial volume fractions of species $\phi_i$ differ. Species 1, 2 and 3 start with an an initial volume fraction of $\phi_{1} = \phi_{2} = \phi_{3} = 0.02$, species 4 starts with $\phi_4 = 0.2$}
     \label{fig: Case2 4Species}
\end{figure}
\begin{figure}[H]
    \centering
    \begin{subfigure}{0.45\textwidth}
        \centering
        \includegraphics[width=1\textwidth]{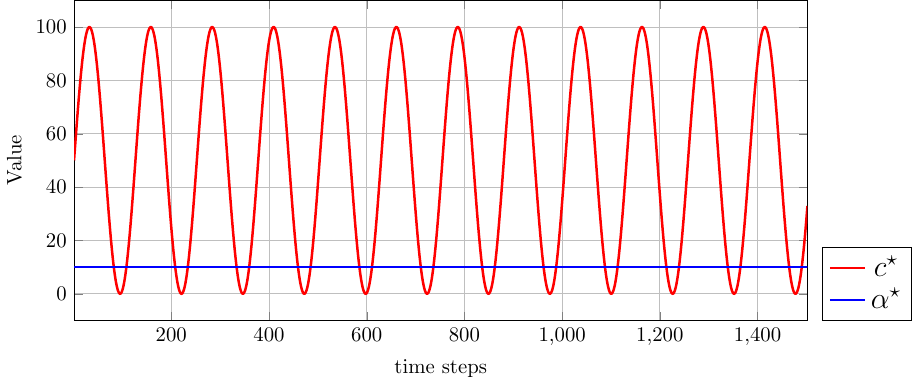}
        \caption{Nutrients and antibiotics for test case 3}
        \label{subfig: Case1b Changing}
    \end{subfigure}
    \begin{subfigure}{0.45\textwidth}
        \centering
        \includegraphics[width=1\textwidth]{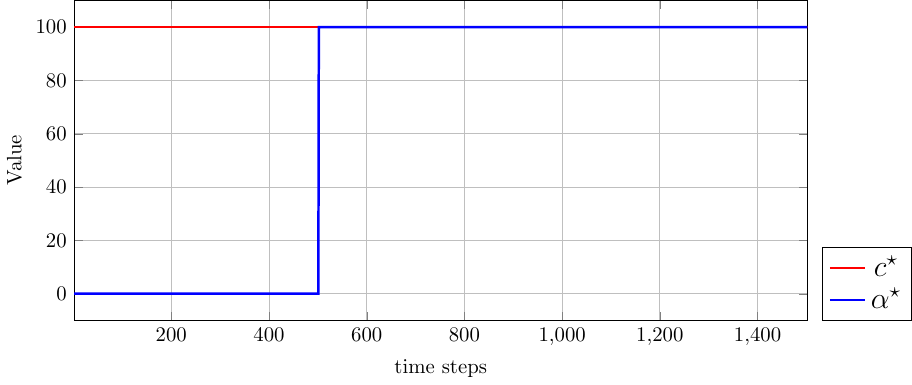}
        \caption{Nutrients and antibiotics for test case 4}
        \label{subfig: Case2b Changing}
     \end{subfigure}   
     \caption{Functions with which nutrients and antibiotics are applied in test cases 3 and 4.}
     \label{fig: Case1b Changing}
\end{figure}

\begin{figure}[H]
    \centering
    \begin{subfigure}{0.45\textwidth}
        \includegraphics[width=1\textwidth]{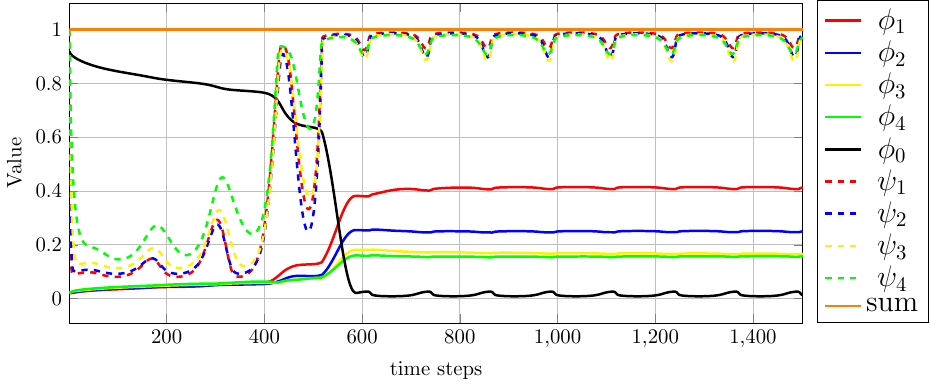}
        \caption{Volume fractions $\phi_i$ and percentage of living cells $\psi_i$ of test case 3}
        \label{subfig: Case1 Changing}
    \end{subfigure}
    \begin{subfigure}{0.45\textwidth}
        \includegraphics[width=1\textwidth]{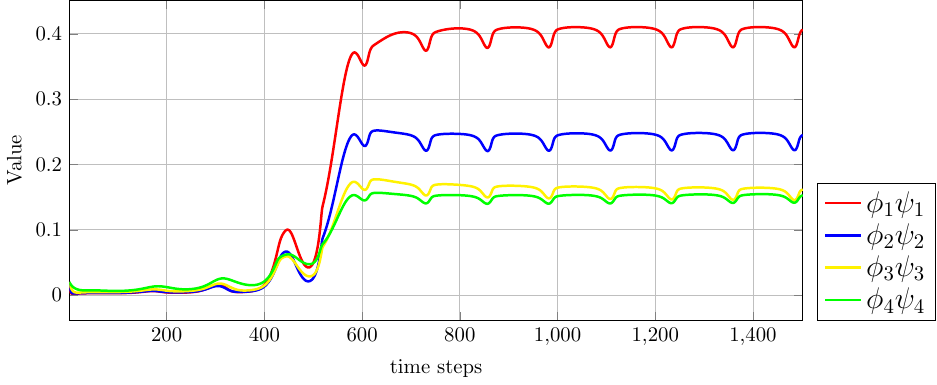}
        \caption{Amount of living cells $\Bar{\phi}_i$ of test case 3\newline}
        \label{subfig: Case1a Changing}
    \end{subfigure}
     \caption{Result of the third test case with four species. Nutrients are applied with a sine function.}
     \label{fig: Case1 Changing}
\end{figure}

\begin{figure}[H]
    \centering
    \begin{subfigure}{0.45\textwidth}
        \includegraphics[width=1\textwidth]{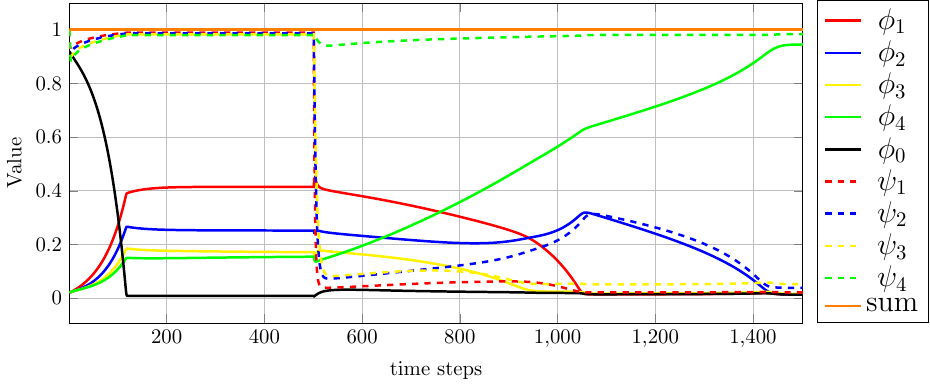}
        \caption{Volume fractions $\phi_i$ and percentage of living cells $\psi_i$ of test case 4}
        \label{subfig: Case2 Changing}
    \end{subfigure}
    \begin{subfigure}{0.45\textwidth}
        \includegraphics[width=1\textwidth]{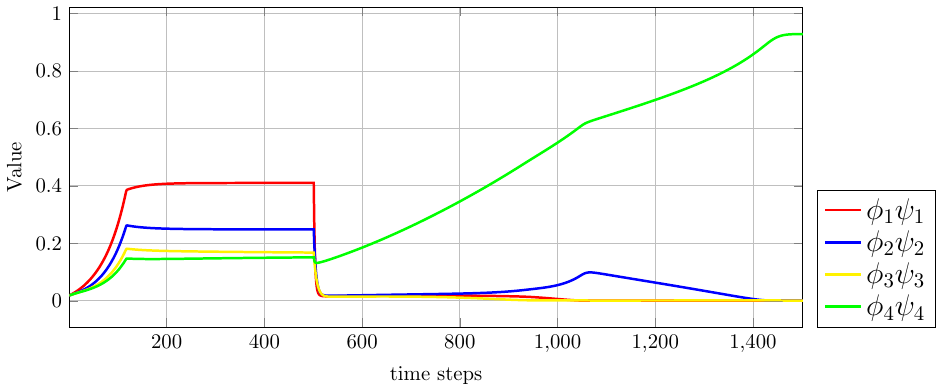}
        \caption{Amount of living cells $\Bar{\phi}_i$ of test case 4\newline}
        \label{subfig: Case2a Changing}
    \end{subfigure}
     \caption{Result of the forth test case with four species. Antibiotics are applied with a step function at time step $500$.}
     \label{fig: Case2 Changing}
\end{figure}

 
\noindent In \autoref{fig: Case1 4Species} and \autoref{fig: Case2 4Species}, the results can be seen. The first case shows all four species with the same initial biofilm concentration. Since the antibiotic sensitivity differs in all four species, with species~1 being the most sensitive and species~4 being the least sensitive, the initial drop in living percentage of each species $\psi_i$ is also different. Species~4 is the only species with a minimum of $20 \%$ of its cells still living. Due to the greater growing parameter of the first species, however, in the course of the simulation, the first species takes over as the dominant species in the biofilm. The different viscosities $\eta_i$ determine how fast each species grows. This can be seen in \autoref{subfig: Case1 4Species} and \autoref{subfig: Case2 4Species}, where the first species' gradient in the beginning is much greater than that of each other species. The second case investigated here only differs in the initial conditions, i.e., the fourth species gets a head start in the simulation. Since there is now less free space available to all species of biofilm, the equilibrium between all species is reached earlier. In addition to that, by the last time step, species 4 still has more volume than species 3, which, in the prior experiment, had more.
\noindent For the nutrients, as well as the antibiotics, an arbitrary function can be implemented. In the following test case, the nutrients were applied with a Sine function (see \autoref{subfig: Case1b Changing}). The results can be seen in \autoref{subfig: Case1 Changing} and \autoref{subfig: Case1a Changing}. The general shape of the results roughly mimics these from \autoref{subfig: Case1 4Species}, with the exception of the fluctuation due to little (or no) nutrients in the system.
\noindent For the final case investigated, the sensitivity towards antibiotics was adjusted. The first species now has a high sensitivity of $b_1 = 10$, while the fourth species is almost resistant against antibiotics with $b_4 = 0.01$. After one-third of the simulation time, antibiotics are inserted into the system (see \autoref{subfig: Case2b Changing}). Every species, except the resistant species 4, almost immediately dies, as can be observed in \autoref{subfig: Case2 Changing} and \autoref{subfig: Case2a Changing}. After the insertion of antibiotics, resistant species 4 takes over the space which was inhabited by the other species prior to the insertion. The second species, which is not as sensitive to antibiotics but has a smaller viscosity than species 4, i.e. $\eta_2 \textcolor{black}{<} \eta_4$, and is thus more reactive to changes in the system, regains some space but is ultimately dominated by the fourth species. In this example, cooperative behavior is not the dominating factor of species survival. The resistance against antibiotics is more important.

\section{Conclusion and outlook}\label{chapt:Conclusion}
In this work, a new multi-species biofilm model is presented. In principle, an arbitrary number of species can be described using this model. For the description, one internal variable is needed to describe the empty space as well as two additional internal variables for each biofilm. In this work, biofilm is modeled with the variables of the volume fraction of all, living and dead, biofilm $\phi$, as well as the percentage of living cells inside the volume fraction $\psi$. This leads to the  amount of living cells given by $\Bar{\phi} = \phi \psi$. The interactions between each species are modeled with energies. For the evolution of biofilm, a dissipated function is introduced as a function of the living bacteria $\Bar{\phi}$ and thus a product of the internal state variables, leading to a coupled system of evolution equations, capable of non-linear effects. The free energy density is modeled with little assumptions concerning the growth and death of biofilm. Nevertheless, the model is capable of producing complex results. Many behaviors as observed in multi-species biofilm can be modeled. The reduced number of parameters of the model simplifies the pending validation of the model with \textit{in vitro} experiments. The model is also suitable for further investigation of biofilm behavior, for example, accounting for stochastic properties.
\section*{Funding and acknowledgment}
The authors highly acknowledge the funding by the German Research Foundation (Deutsche Forschungsgemeinschaft, DFG) through the project grant SFB/TRR-298-SIIRI – Project-ID 426335750. 
\\\\
Funded by the European Union (ERC, Gen-TSM, project number 101124463). Views and opinions expressed are however those of the author(s) only and do not necessarily reflect those of the European Union or the European Research Council Executive Agency. Neither the European Union nor the granting authority can be held responsible for them. 

\section*{Data availability statement}
The datasets used and/or analysed during the current study is available from the corresponding author on reasonable request.

\section*{Declaration of interests}
The authors declare that they have no known competing financial interests or personal relationships that could have appeared to influence the work reported in this paper.

\appendix
\section{Appendix A}\label{appendix:A}
\textcolor{black}{Additional simulation were performed to further investigate the behavior of the viscosities $\eta_i$. The simulations provided here are based on the cases 4 and 5 for the two species model respectively.}
\begin{centering}
    \begin{table*}[h!]
\begin{tabular}{llrrrr}\toprule
\textbf{variable} & \textbf{unit} & \textbf{Case 4${\star}$} & \textbf{Case 5${\star}$} & \textbf{Case 4${\star\star}$} & \textbf{Case 5${\star\star}$} \\ \midrule
${a}_{11}$ & [-] & 1 & 1 & 1 & 1  \\ \hdashline
${a}_{12}$ & [-] & 0 & 0 & 0 & 0 \\ \hdashline
${a}_{22}$ & [-] & 1 & 1 & 1 & 1\\ \midrule

${b}_{1}$ & [-] & 1 & 1 & 1 & 1  \\ \hdashline
${b}_{2}$ & [-] & 2 & 2 & 2 & 2 \\ \midrule

${\eta}_{1}$ & $[\si[per-mode=fraction]{\kg\per\meter\per \time }]$ & 2 & 2 & 1 & 1 \\ \hdashline
${\eta}_{2}$ & $[\si[per-mode=fraction]{\kg\per\meter\per\time}]$ & 1 & 1 & 1 & 1 \\ \midrule

{initial $\phi_{1}$} & [-] & 0.2 & 0.25 & 0.2 & 0.25\\ \hdashline
{initial $\phi_{2}$} & [-] & 0.3 & 0.3 & 0.3 & 0.3\\ \bottomrule

\end{tabular}
\caption{Values for the parameters of the additional simulations performed with two species present. The variables describing the nutrients $\mathrm{c}^{\star}$ and the antibiotics $\alpha^{\star}$ are left constant at $\mathrm{c}^{\star} = \SI[per-mode=fraction]{100}{\joule\per\meter\cubed}$ and $\alpha^{\star} = \SI[per-mode=fraction]{10}{\joule\per\meter\cubed}$.}
\label{tab:appendixA}
\end{table*}
\end{centering}

\begin{figure}[H]
    \centering
    \begin{subfigure}{0.45\textwidth}
        \centering
        \includegraphics[width=1\textwidth]{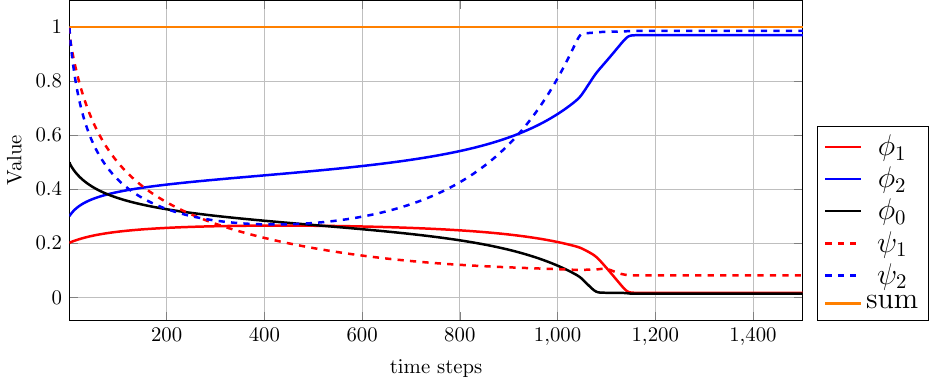}
        \caption{Volume fractions $\phi_i$ and percentage of living cells $\psi_i$ of test case 4$\star$}
        \label{subfig: appendixA1}
    \end{subfigure}
    \begin{subfigure}{0.45\textwidth}
        \centering
        \includegraphics[width=1\textwidth]{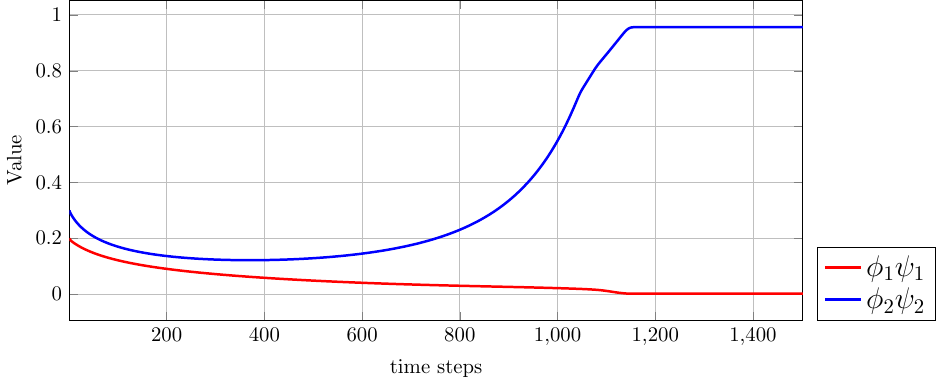}
        \caption{Amount of living cells $\Bar{\phi}_i$ of test case 4$\star$\newline}
        \label{subfig: appendixA1a}
    \end{subfigure}  
     \caption{Results of the additional case with two species based on test case 4 with varying viscosites. The viscosities have been switched compared to case 4.}
     \label{fig: appendixA1}
\end{figure} 

\begin{figure}[H]
    \centering
    \begin{subfigure}{0.45\textwidth}
        \centering
        \includegraphics[width=1\textwidth]{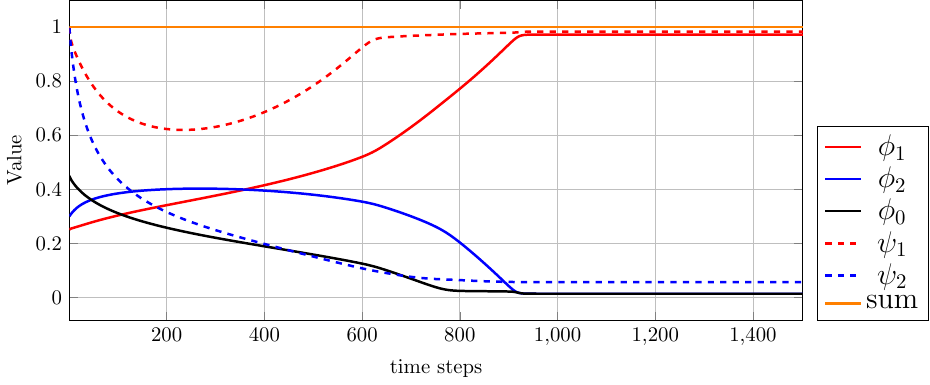}
        \caption{Volume fractions $\phi_i$ and percentage of living cells $\psi_i$ of test case 5$\star$}
        \label{subfig: appendixA2}
    \end{subfigure}
    \begin{subfigure}{0.45\textwidth}
        \centering
        \includegraphics[width=1\textwidth]{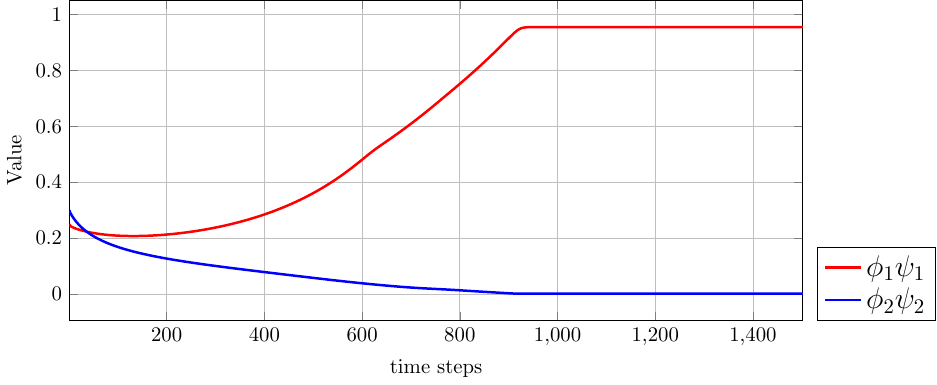}
        \caption{Amount of living cells $\Bar{\phi}_i$ of test case 5$\star$\newline}
        \label{subfig: appendixA2a}
    \end{subfigure}  
     \caption{Results of the additional case with two species based on test case 5 with varying viscosites. The viscosities have been switched compared to case 5.}
     \label{fig: appendixA2}
\end{figure} 

\begin{figure}[H]
    \centering
    \begin{subfigure}{0.45\textwidth}
        \centering
        \includegraphics[width=1\textwidth]{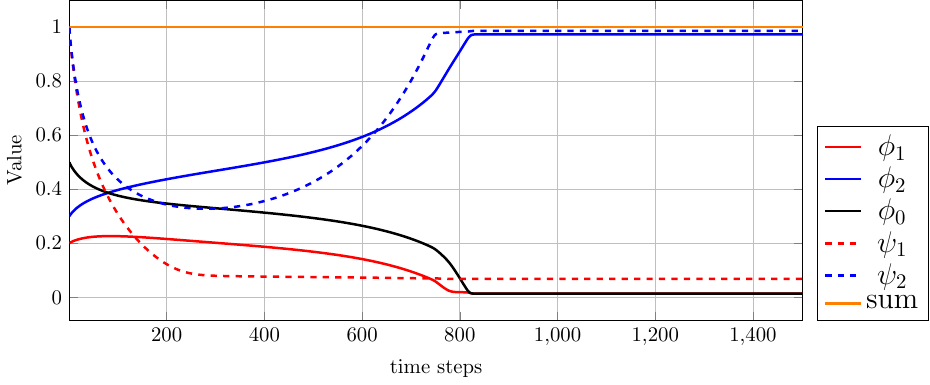}
        \caption{Volume fractions $\phi_i$ and percentage of living cells $\psi_i$ of test case 4$\star\star$}
        \label{subfig: appendixA3}
    \end{subfigure}
    \begin{subfigure}{0.45\textwidth}
        \centering
        \includegraphics[width=1\textwidth]{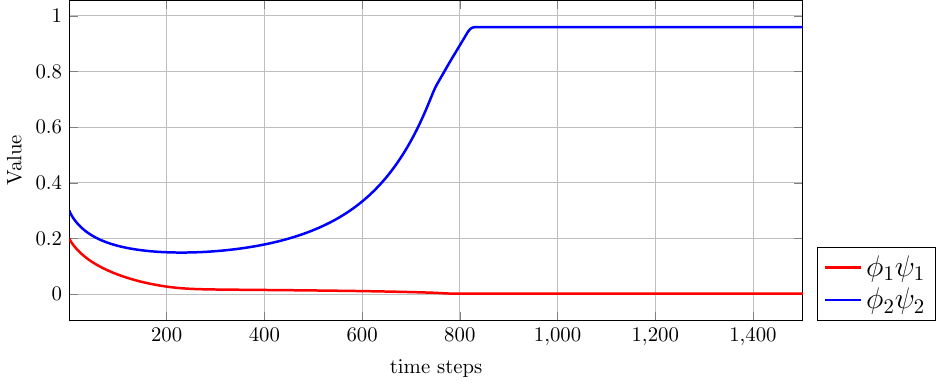}
        \caption{Amount of living cells $\Bar{\phi}_i$ of test case 4$\star\star$\newline}
        \label{subfig: appendixA3a}
    \end{subfigure}  
     \caption{Results of the additional case with two species based on test case 4 with varying viscosites. The viscosities have been set equal.}
     \label{fig: appendixA3}
\end{figure} 

\begin{figure}[H]
    \centering
    \begin{subfigure}{0.45\textwidth}
        \centering
        \includegraphics[width=1\textwidth]{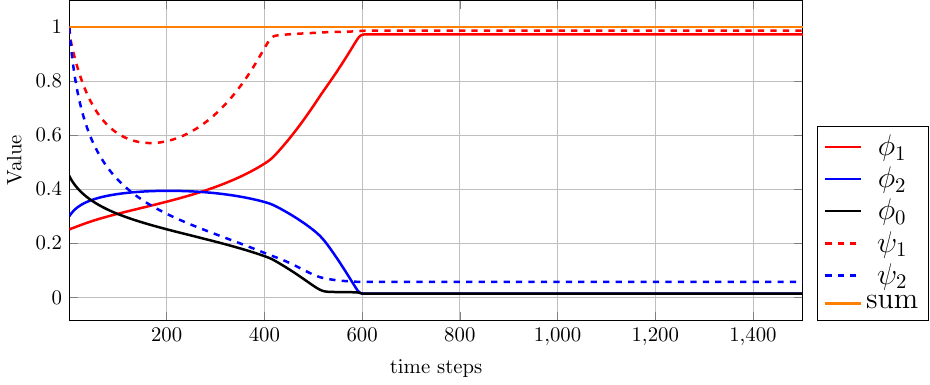}
        \caption{Volume fractions $\phi_i$ and percentage of living cells $\psi_i$ of test case 5$\star\star$}
        \label{subfig: appendixA4}
    \end{subfigure}
    \begin{subfigure}{0.45\textwidth}
        \centering
        \includegraphics[width=1\textwidth]{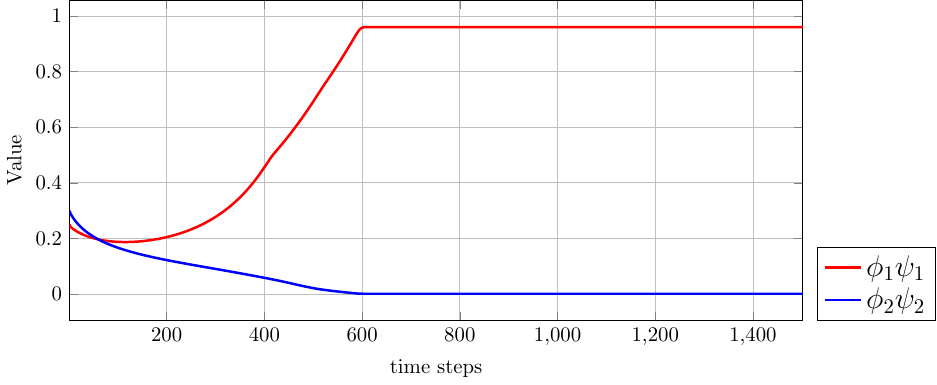}
        \caption{Amount of living cells $\Bar{\phi}_i$ of test case 5$\star\star$\newline}
        \label{subfig: appendixA4a}
    \end{subfigure}  
     \caption{Results of the additional case with two species based on test case 5 with varying viscosites. The viscosities have been set equal.}
     \label{fig: appendixA4}
\end{figure} 

\addcontentsline{toc}{chapter}{Bibliography}
\bibliographystyle{plain}
\bibliography{bib}

\end{document}